\documentclass[dvips]{article}
\usepackage{supertabular,lscape,epsfig}
\usepackage{amssymb}
\usepackage{amsmath}
\DeclareSymbolFont{ppa}{OT1}{ppl}{m}{it}
\DeclareMathSymbol{\vv}{\mathalpha}{ppa}{'166}

\DeclareSymbolFont{ppa}{OT1}{ppl}{m}{it}
\DeclareMathSymbol{\vv}{\mathalpha}{ppa}{'166}

\begin{document}

\newcommand{\dd}{\,{\rm d}}
\newcommand{\ie}{{\it i.e.},\,}
\newcommand{\etal}{{\it et al.\ }}
\newcommand{\eg}{{\it e.g.},\,}
\newcommand{\cf}{{\it cf.\ }}
\newcommand{\vs}{{\it vs.\ }}
\newcommand{\zdot}{\makebox[0pt][l]{.}}
\newcommand{\up}[1]{\ifmmode^{\rm #1}\else$^{\rm #1}$\fi}
\newcommand{\dn}[1]{\ifmmode_{\rm #1}\else$_{\rm #1}$\fi}
\newcommand{\upd}{\up{d}}
\newcommand{\uph}{\up{h}}
\newcommand{\upm}{\up{m}}
\newcommand{\ups}{\up{s}}
\newcommand{\arcd}{\ifmmode^{\circ}\else$^{\circ}$\fi}
\newcommand{\arcm}{\ifmmode{'}\else$'$\fi}
\newcommand{\arcs}{\ifmmode{''}\else$''$\fi}
\newcommand{\MS}{{\rm M}\ifmmode_{\odot}\else$_{\odot}$\fi}
\newcommand{\RS}{{\rm R}\ifmmode_{\odot}\else$_{\odot}$\fi}
\newcommand{\LS}{{\rm L}\ifmmode_{\odot}\else$_{\odot}$\fi}

\newcommand{\Abstract}[2]{{\footnotesize\begin{center}ABSTRACT\end{center}
\vspace{1mm}\par#1\par
\noindent
{~}{\it #2}}}

\newcommand{\TabCap}[2]{\begin{center}\parbox[t]{#1}{\begin{center}
  \small {\spaceskip 2pt plus 1pt minus 1pt T a b l e}
  \refstepcounter{table}\thetable \\[2mm]
  \footnotesize #2 \end{center}}\end{center}}

\newcommand{\TableSep}[2]{\begin{table}[p]\vspace{#1}
\TabCap{#2}\end{table}}

\newcommand{\FigCap}[1]{\footnotesize\par\noindent Fig.\  %
  \refstepcounter{figure}\thefigure. #1\par}

\newcommand{\TableFont}{\footnotesize}
\newcommand{\TableFontIt}{\ttit}
\newcommand{\SetTableFont}[1]{\renewcommand{\TableFont}{#1}}

\newcommand{\MakeTable}[4]{\begin{table}[htb]\TabCap{#2}{#3}
  \begin{center} \TableFont \begin{tabular}{#1} #4 
  \end{tabular}\end{center}\end{table}}

\newcommand{\MakeTableSep}[4]{\begin{table}[p]\TabCap{#2}{#3}
  \begin{center} \TableFont \begin{tabular}{#1} #4 
  \end{tabular}\end{center}\end{table}}
\newcommand{\TabCapp}[2]{\begin{center}\parbox[t]{#1}{\centerline{
  \small {\spaceskip 2pt plus 1pt minus 1pt T a b l e}
  \refstepcounter{table}\thetable}
  \vskip2mm
  \centerline{\footnotesize #2}}
  \vskip3mm
\end{center}}

\newcommand{\MakeTableSepp}[4]{\begin{table}[p]\TabCapp{#2}{#3}\vspace*{-.7cm}
  \begin{center} \TableFont \begin{tabular}{#1} #4 
  \end{tabular}\end{center}\end{table}}

\newfont{\bb}{ptmbi8t at 12pt}
\newfont{\bbb}{cmbxti10}
\newfont{\bbbb}{cmbxti10 at 9pt}
\newcommand{\uprule}{\rule{0pt}{2.5ex}}
\newcommand{\douprule}{\rule[-2ex]{0pt}{4.5ex}}
\newcommand{\dorule}{\rule[-2ex]{0pt}{2ex}}
\def\thefootnote{\fnsymbol{footnote}}

\newenvironment{references}%
{
\footnotesize \frenchspacing
\renewcommand{\thesection}{}
\renewcommand{\in}{{\rm in }}
\renewcommand{\AA}{Astron.\ Astrophys.}
\newcommand{\AAS}{Astron.~Astrophys.~Suppl.~Ser.}
\newcommand{\ApJ}{Astrophys.\ J.}
\newcommand{\ApJS}{Astrophys.\ J.~Suppl.~Ser.}
\newcommand{\ApJL}{Astrophys.\ J.~Letters}
\newcommand{\AJ}{Astron.\ J.}
\newcommand{\IBVS}{IBVS}
\newcommand{\PASP}{P.A.S.P.}
\newcommand{\Acta}{Acta Astron.}
\newcommand{\MNRAS}{MNRAS}
\renewcommand{\and}{{\rm and }}
\section{{\rm REFERENCES}}
\sloppy \hyphenpenalty10000
\begin{list}{}{\leftmargin1cm\listparindent-1cm
\itemindent\listparindent\parsep0pt\itemsep0pt}}%
{\end{list}\vspace{2mm}}

\def\TYLDA{~}
\newlength{\DW}
\settowidth{\DW}{0}
\newcommand{\dw}{\hspace{\DW}}

\newcommand{\refitem}[5]{\item[]{#1} #2%
\def\REFARG{#3}\ifx\REFARG\TYLDA\else, {\it#3}\fi
\def\REFARG{#4}\ifx\REFARG\TYLDA\else, {\bf#4}\fi
\def\REFARG{#5}\ifx\REFARG\TYLDA\else, {#5}\fi.}

\newcommand{\Section}[1]{\section{\hskip-6mm.\hskip3mm#1}}
\newcommand{\Subsection}[1]{\subsection{#1}}
\newcommand{\Acknow}[1]{\par\vspace{5mm}{\bf Acknowledgements.} #1}
\pagestyle{myheadings}

\newcommand{\xrule}{\rule{0pt}{2.5ex}}
\newcommand{\xxrule}{\rule[-1.8ex]{0pt}{4.5ex}}
\def\thefootnote{\fnsymbol{footnote}}
\begin{center}
{\Large\bf The Optical Gravitational Lensing Experiment.\\
\vskip3pt
Small Amplitude Variable Red Giants\\
\vskip6pt
in the Magellanic Clouds\footnote{Based on observations obtained with the
1.3~m Warsaw telescope at the Las Campanas Observatory of the Carnegie
Institution of Washington.}}
\vskip1.2cm
{\bf I.~~S~o~s~z~y~\'n~s~k~i$^{1,2}$,~~A.~~U~d~a~l~s~k~i$^1$,~~M.~~K~u~b~i~a~k$^1$,\\
M.~~S~z~y~m~a~{\'n}~s~k~i$^1$,~~G.~~P~i~e~t~r~z~y~\'n~s~k~i$^{1,2}$,~~K.~\.Z~e~b~r~u~\'n$^1$,\\
~~O.~~S~z~e~w~c~z~y~k$^1$~ and ~\L.~~W~y~r~z~y~k~o~w~s~k~i$^1$}
\vskip8mm
{$^1$Warsaw University Observatory, Al.~Ujazdowskie~4, 00-478~Warszawa,
Poland\\
e-mail: (soszynsk,udalski,msz,mk,pietrzyn,zebrun,szewczyk,wyrzykow)@astrouw.edu.pl\\
$^2$ Universidad de Concepci{\'o}n, Departamento de Fisica,
Casilla 160--C, Concepci{\'o}n, Chile}
\end{center}

\vskip0.6cm 

\Abstract{We present analysis of the large sample of variable red giants
from the Large and Small Magellanic Clouds detected during the second
phase of the Optical Gravitational Lensing Experiment (OGLE-II) and
supplemented with OGLE-III photometry. Comparing pulsation properties
of detected objects we find that they constitute two groups with clearly
distinct features. In this paper we analyze in detail small amplitude
variable red giants (about 15\,400 and 3000 objects in the LMC and SMC,
respectively). The vast majority of these objects are multi-periodic. At
least 30\% of them exhibit two modes closely spaced in the power
spectrum, what likely indicates non-radial oscillations. About 50\%
exhibit additional so called Long Secondary Period.

To distinguish between AGB and RGB red giants we compare {\it PL}
diagrams of multi-periodic red giants located above and below the tip of
the Red Giant Branch (TRGB).  The giants above the TRGB form four
parallel ridges in the {\it PL} diagram. Among much more numerous sample
of giants below the TRGB we find objects located on the low luminosity
extensions of these ridges, but most of the stars are located on the
ridges slightly shifted in $\log P$. We interpret the former as the
second ascent AGB red giants and the latter as the first ascent RGB
objects. Thus, we empirically show that the pulsating red giants fainter
than the TRGB are a mixture of RGB and AGB giants.

Finally, we compare the Petersen diagrams of the LMC, SMC and Galactic
bulge variable red giants and find that they are basically identical
indicating that the variable red giants in all these different stellar
environments share similar pulsation properties.}{{\bf Key words:} Stars: oscillations --
Stars: late-type -- Stars: AGB and post-AGB -- Magellanic Clouds}

\Section{Introduction}
In recent years a significant progress has been made in studies of
pulsations of red giants. The major breakthrough occurred when the large
microlensing surveys data of huge samples of these stars spanning several
years and all-sky IR surveys photometry became available.

Discovered years ago near-infrared period-luminosity ({\it PL}) relation
for Mira-type stars (\eg Glass and Lloyd Evans 1981) was extended to
semi-regular variables by Wood and Sebo (1996) who showed the second
sequence in the $\log{P}$--$K$ diagram of the long period variables
(LPV). Subsequently, Wood \etal (1999) analyzing MACHO photometric data of
stars in the LMC revolutionized this picture showing five parallel ridges
in the {\it PL} diagram (denoted as A--E). These results were later
confirmed by independent studies based on data originated in various
projects: EROS+DENIS (Cioni \etal 2001), MOA (Noda \etal 2002),
AGAPEROS+DENIS (Lebzelter, Schultheis and Melchior 2002).

Kiss and Bedding (2003, 2004) used photometric data from the OGLE-II
catalog of variable stars ({\.Z}ebru{\'n} \etal 2001) supplemented with
{\it K}-band photometry from 2MASS survey to analyze variable red giants in
the Magellanic Clouds. They discovered separate {\it PL} relations of a
huge number of variables above and below the tip of the red giant branch
(TRGB). They distinguished four ridges in the $\log{P}$--$K$ plane of the
pulsating AGB stars and three sequences of giants fainter than the
TRGB. They argued that a substantial fraction of variable giants below the
TRGB are in the RGB phase.

Ita \etal (2004) presented basically identical, albeit sharper picture of
the {\it K}-band {\it PL} relation using OGLE-II data and more accurate
{\it K}-band photometry obtained from their SIRIUS IR survey of the
Magellanic Clouds. OGLE-II variable red giants were also analyzed by
Groenewegen (2004).

Small amplitude variability of red giants has been known for decades.
Stebbins and Huffer (1930) surveyed photoelectrically red giants and
detected variability of many stars of type M0 and later. They discovered
that the cooler the star, the larger the scatter of magnitudes. In
recent years this rule was extended to K giants. Edmonds and Gilliland
(1996) analyzed HST observations of K giants in the globular cluster
47~Tuc. They found variability with periods of 2--4~days and
semi-amplitudes of about 5--15~mmag in K3 and later giants. Other
surveys of K giants (Jorissen \etal 1997, Henry \etal 2000) confirmed
that microvariability of these stars is a common feature. Tens of
Galactic small amplitude red variable stars were monitored in long term
monitoring programs by John Percy (\eg Percy, Wilson and Henry 2001).

Small amplitude red giants were also detected among huge number of
variables in the Galactic bulge. Wray, Eyer and Paczy{\'n}ski (2004)
analyzed the periodicity of 200\,000 objects from the OGLE-II catalog of
variable stars in the Galactic bulge (Wo{\'z}niak \etal 2002). They
selected 18,000 stars (calling them OGLE Small Amplitude Red Giants --
OSARGs) with periods ${10<P<100}$~days and {\it I}-band amplitudes in the
range ${0.005<A<0.13}$~mag. The variables clearly followed two distinct
period--amplitude relations.

The nature of small amplitude pulsations of red giants is still not
clear. Two mechanisms are proposed: self excitations of unstable modes (so
called Mira-like pulsations) and convection induced excitation of linearly
stable modes (solar-like oscillations). Both possibilities were discussed
by Dziembowski \etal (2001) and Christensen-Dalsgaard \etal (2001).

One of the by-products of the OGLE survey is a huge number of variable
stars found in the Magellanic Clouds. The main purpose of this paper is to
provide possibly most precise characteristics of poorly known type of
pulsating stars -- OGLE Small Amplitude Red Giants in the Large and Small
Magellanic Clouds.

\vspace*{0.5cm}
\Section{Observations and Data Reductions}
All observations presented in this paper were carried out during the second
and the third phases of the OGLE experiment with the 1.3-m Warsaw telescope
at Las Campanas Observatory, Chile. The observatory is operated by the
Carnegie Institution of Washington. The OGLE-II project started in January
1997. The telescope was equipped with the ``first generation'' camera with
a~SITe ${2048\times2048}$ CCD detector working in drift-scan mode. The
pixel size was 24~$\mu$m giving the 0.417~arcsec/pixel scale. Observations
were performed in the ``slow'' reading mode of the CCD detector with the
gain 3.8~e$^-$/ADU and readout noise of about 5.4~e$^-$. For details of the
instrumentation setup of OGLE-II, we refer the reader to Udalski, Kubiak
and Szyma{\'n}ski (1997). The OGLE-II fields cover 4.5 square degrees in
the central parts of the LMC and 2.4 square degrees in the SMC.

Second phase of the OGLE project was completed in November 2000 and
in June 2001 the third stage of the experiment began. The Warsaw
telescope was equipped with a ``second generation'' CCD mosaic camera
consisting of eight SITe ST-002a CCD detectors with ${2048\times4096}$
pixels of 15~$\mu$m size (Udalski 2003a). This corresponds to
0.26~arcsec/pixel scale and the field of view of the whole mosaic
$35\arcm\times35\arcm$. The last observations presented in this paper
were collected in November 2003, so the analyzed data span 8~years.

Photometry was obtained with {\it I} and {\it V} filters, closely
resembling the standard system. For the period analysis {\it I}-band
observations were used, in which the majority of frames were taken. The
OGLE-II photometry was obtained using the Difference Image Analysis (DIA)
method -- image subtraction algorithm developed by Alard and Lupton (1998)
and Alard (2000), and implemented by Wo{\'z}niak (2000). Contrary to the
general catalog of variable stars in the Magellanic Clouds ({\.Z}ebru{\'n}
\etal 2001), the DIA photometry was reprocessed for all stars found in the
reference images of the fields. In this way more complete sample, in
particular of very small amplitude variable red giants, could be detected
among monitored stars. The OGLE-III photometry comes from the standard
OGLE-III photometry pipeline (Udalski 2003a) and is also based on DIA.

We tied photometry obtained during the second and third phases of the OGLE
survey by shifting the OGLE-III magnitudes to well calibrated OGLE-II
photometry. For each object we determined the median difference between
OGLE-III and OGLE-II luminosities of at least several dozen constant stars
in the closest neighborhood. This value was then used as the correction of
the OGLE-III magnitudes. Combined OGLE-II and OGLE-III datasets with its
much longer time span made it possible to derive much more complete and
accurate periodicities by filtering out spurious, non-coherent and unstable
frequencies possible in the OGLE-II dataset alone.

Altogether, we collected from about 400 to 800 observations in {\it I}
filter (depending on the field) and about 30--70 measurements in the {\it
V}-band. In the OGLE-II phase the effective exposure time lasted 125 and
174~seconds for the {\it I} and {\it V}-band, respectively. OGLE-III
exposure time of observations in {\it I}-band was increased to 180
seconds. The median seeing of our dataset was about 1\zdot\arcs3.

In the majority of other analysis of red giant pulsations near infrared
photometric bands were used, especially {\it K}-band magnitudes. Stars in
the latest phases of evolution are characterized by the high mass-loss, and
they are often obscured by significant amount of dust. Therefore, infrared
bands minimize the scatter of the {\it PL} sequences caused by interstellar
extinction. On the other hand infrared magnitudes usually come from single
epoch measurements what increases the scatter.

We used in our analysis reddening free Wesenheit index, $W_I$, defined as:
$$W_I=I-1.55(V-I)$$ 
where {\it I} and {\it V} are intensity mean magnitudes and 1.55 is the
mean ratio of total-to-selective absorption ($A_I/E(V-I)$). It is worth
noticing that the {\it PL} diagrams which use $W_I$ index present sequences
fairly narrow, similar or even better defined than in the $\log{P}$--$K$
diagrams.

\vspace*{0.5cm}
\Section{Selection of OSARGs}
We searched for variable red giants among {\it I}-band light curves of
all stars brighter than 17~mag and 17.5~mag in the LMC and SMC,
respectively. Usually, a search for variable stars is preceded by a
preliminary selection based, for instance, on scatter of observations.
However, we performed the period analysis of each star, because
substantial number of small amplitude variable red giants could have
been rejected during the preselection process.

At the first stage of the search we performed a low-resolution period
analysis of about 260\,000 and 120\,000 objects in the LMC and SMC,
respectively. We used program {\sc Fnpeaks} (Ko{\l}aczkowski 2003, private
communication) which implements the algorithm of Discrete Fourier
Transform.

After selection of the highest peak in the power spectrum corresponding
to the primary dominant period we examined periodograms with higher
resolution and determined more accurate primary periods. Then, the light
curve of analyzed object was folded with derived period, approximated by
third order Fourier series and fitted functions were subtracted from the
observational data. The residuals were again searched for other periodic
signal by repeating the procedure. Only periods with the signal-to-noise
parameter larger than 3.6 were considered as real. Up to five  dominant
periods per object were kept. Finally, all objects with no {\it V}-band
measurements and stars with colors ${V-I<1.0}$ (\eg Cepheids, main
sequence stars) were rejected.

The input list of stars consisted of both, variable and non-variable,
objects. Separation of red giants with the smallest amplitudes from
non-variable stars turned out to be quite difficult. In some cases neither
the scatter of the measurements, nor signal-to-noise parameter produced by
the period searching program, nor even visual inspection of the light
curves, allowed to distinguish between variables and
non-variables. Therefore, for each object we performed a second periodicity
search using another program, {\sc Predator}, (Mizerski 2004, private
communication) based on another method of the frequency analysis
(Lomb-Scargle algorithm). For further studies we limited our sample to
stars which the dominant period derived during the first search was found
between 25 highest peaks of the periodogram obtained in the second
stage. This method allowed us to reject the vast majority of non-variable
objects.

The vast majority of the selected small amplitude variables are
multi-periodic. Certainly, a small part of periods derived in stars from
our sample might be spurious. This is inevitable when analyzing light
curves of variables with so small amplitudes. However, the visual
inspection of light curves indicates, that the vast majority of
variables have dominant periods well determined, what combined with
large number of objects in the final sample enabled us to study
statistical features of variable red giants.

\begin{figure}[htb]
\vspace{8.5cm}
%\centerline{\includegraphics[width=13cm, bb=35 395 570 750]{fig1.ps}} 
\FigCap{Period--$W_I$ diagrams for variable red giants in the LMC. The
upper diagrams are constructed using primary periods with ${S/N>6}$. The
lower diagrams show the second and third dominant periods--$W_I$ relation
for giants from the ridge A ({\it left panel}) and C ({\it right panel}).}
\end{figure} 

In the upper panels in Fig.~1 we show the {\it PL} relation,
$\log{P}$--$W_I$, for our sample of variable red giants in the LMC. Only
primary periods with signal-to-noise parameter larger than 6 are
presented: altogether 16\,000 stars. {\it PL} diagram  showing several
clear well separated sequences of pulsating red giants  is basically
identical with infrared diagram for the {\it K}-band (Ita \etal 2004).
We labeled the ridges  A--D following notation of Wood \etal (1999),
adding label C$''$ for a clear sequence between C and D, weakly seen in
the {\it K}-band (Ita \etal 2004). The sequence denoted as C$'$ by Ita
\etal (2004) is practically merged with the ridge B in Fig.~1.

Lower panels in Fig.~1 show the {\it PL} relations constructed using the
second and third dominant periods of stars highlighted in magenta in the
upper panels, that is the sequence A (3800 objects) in the left diagram,
and sequence C (1400 objects) in the right diagram. As can be noticed,
both groups of red giants form completely different patterns in the
secondary period--$W_I$ plane. Stars with the primary periods in the
sequence A reveal four narrow {\it PL} ridges corresponding to the
sequences A and B, and additional wider sequence corresponding to the
sequence D. Sequences C and  C$''$ are not present at all in the
bottom-left plot in Fig.~1. 

On the other hand, stars with the primary periods located in the
sequence C also form clear ridges in the bottom-right panel of Fig.~1.
They correspond to the sequences B, C and less clearly C$''$. The ridge
D is practically not seen in this panel. It is also evident that none
of the secondary periods of this group correspond to the sequence~A.

This striking difference between variable giants belonging to ridges
A and C suggests that both groups represent different types of pulsating
stars. The first group is referred in this paper as OSARG. The second
group of long period variables (LPV), consisting of semi-regular
variables (SRV) and Mira type variables, will be analyzed in a
forthcoming paper. It is worth noticing here that in both groups of
stars many secondary periodicities are very close to the primary ones
(\ie stars with primary periods in the ridge A or C have other dominant
periods in the ridge A or C, respectively, as well) suggesting
non-radial pulsations of these objects (see Section~4 and 5). 

Sequence B (2500 objects) in the upper panels of Fig.~1 must contain
both types of pulsating giants: OSARG and SRV/Mira stars because this
ridge is populated in both bottom panels of Fig.~1. To separate these
groups we used the second and next (if present) dominant periods. If one
of these periods fell into the sequence A, we classified such an object
as OSARG. Otherwise, the star was added to the list of SRV/Mira objects.

\begin{figure}[p]
\vglue-12mm
\vspace{19cm}
%\centerline{\includegraphics[width=14cm, height=21cm]{fig2.ps}} 
\vspace*{-13mm}
\FigCap{The top panels present the primary period--$W_I$ relation for
variable red giants in the LMC (Fig.~1). The next rows show as follows:
secondary periods--$W_I$ relation, period--amplitude ({\it I}-band)
relation and period--color (${V-I}$) relation for stars highlighted in
magenta in the top panels. Least square fits for period--amplitude and
period--color relation of stars in the left column are repeated for
comparison in  appropriate panels of the right column.}
\end{figure} 

Left panels in Fig.~2 show properties of OSARGs with the primary periods
in the ridge B. Right panels present the same diagrams for SRV/Mira
variables. In the top row {\it PL} relations of the primary dominant
periods are repeated with OSARGs (left) or SRV/Mira stars (right)
highlighted in magenta.

In the next row of Fig.~2 {\it PL} diagrams for other dominant periods
are presented. It is clearly seen that the secondary periods of SRV/Mira
objects populate mostly sequences B (its part denoted as C$'$ by Ita
\etal 2004) and  C similar to objects with the primary periods in the
ridge C (right panels in Fig.~1). On the other hand the secondary
periods of stars from the ridge B classified as {\nobreak OSARGs} closely follow
the pattern of stars with primary periods in the sequence A (left panels
in Fig.~1). Thus, we indeed see that the ridge B for primary periods is
a mixture of OSARG and SRV/Mira variables -- stars with different
pulsational properties. Our criterion of classification seems to
properly separate stars belonging to both groups. 

\begin{figure}[htb]
\vspace{11cm}
%\centerline{\includegraphics[width=11cm, bb=40 225 550 725]{fig3.ps}} 
\vspace*{-3mm}
\FigCap{Color-magnitude diagram of the LMC red giants. Objects classified
as OSARGs are  plotted in magenta and SRV/Mira stars in cyan.}
\end{figure} 
\begin{figure}[p]
\vspace{-1cm} 
\vspace{19cm}
%\centerline{\includegraphics[width=14cm,height=21cm]{fig4.ps}}
\vspace*{-1.3cm}
\FigCap{Examples of the light curves of the OSARG  objects in the LMC.
In the first three upper rows light curves of OSARGs fainter than the
TRGB are presented. In the next two rows examples of OSARGs above the
TRGB are shown. Bottom two rows present variables with distinct long
secondary periods.}
\end{figure} 
The period--amplitude and period--color diagrams for stars with primary
period in the ridge B, presented in the consecutive rows in Fig.~2, reveal
that though the ranges of the amplitudes and colors of both groups of
variables overlap, they follow distinct $\log{P}$--$A$ and
$\log{P}$--$(V-I)$ relations. The SRV/Mira objects typically have larger
amplitudes and span wider range of colors. All these features also suggest
that OSARG and SRV/Mira variables are different types of pulsating stars.

Similar test performed on variable red giants that have the primary
periods in the ridge D in Fig.~1 indicates that this group is also a
mixture of OSARG and SRV/Mira stars, \ie the former have dominant
periodicities in the ridge A while the latter do not.

Fig.~3 presents the  color--magnitude diagram of the LMC stars. Only the
upper part of the red giant branch is shown. Objects classified as OSARG
and SRV/Mira are  plotted in magenta and cyan, respectively.

Fig.~4 shows eight typical light curves of OSARGs in the LMC. The original
data (left panel) and folded light curves (right panel) are arranged
according to periods. To compare amplitudes of variability all diagrams
have the same magnitude scale. Three upper light curves present typical
OSARGs below the TRGB, next two rows show giants brighter than the TRGB,
two lowest light curves are examples of stars with long secondary periods
(see below).

\vspace*{0.5cm}
\Section{OSARGs above the TRGB}
Kiss and Bedding (2003) found a clear drop of stellar density of variable
red giants above the TRGB. They also noticed that {\it PL} relations of
giants brighter and fainter than the TRGB are shifted by about
${\Delta\log{P}\approx0.05}$. Therefore, we separated our sample of OSARGs
into two groups, namely the stars above and below the TRGB and analyzed
them separately. Detailed studies of the multi-periodic stars revealed
significant differences between both groups. Hereafter we refer OSARGs
brighter than the TRGB as type ``a'' (for ``above TRGB'') with the sequence
number (\eg ${\rm a}_1$ means the sequence with the longest
periods). Objects fainter than the TRGB will be described by letter ``b''
(for ``below TRGB'') with the appropriate sequence number.

\begin{figure}[htb]
\vspace{12.9cm}
%\centerline{\includegraphics[width=12.5cm]{fig5.ps}} 
\FigCap{Period-$W_I$ diagrams for OSARG stars brighter than the TRGB
in the LMC (upper panel) and SMC (lower panel). Up to four dominant
periods for each star are plotted.}
\end{figure} 
We used {\it I}-band magnitudes of the TRGB determined by Udalski (2000):
$I_{\rm TRGB}\\ =14.56$~mag in the LMC, and ${I_{\rm TRGB}=14.95}$~mag in
the SMC to separate the stars. Fig.~5 shows the $\log{P}$--$W_I$ {\it PL}
relation of the OSARGs above the TRGB in the LMC and SMC (about 2000 and
400 objects, respectively). For each star up to four points representing
the primary and next three dominant periods are marked. Four distinct
sequences, labeled ${{\rm a}_1-{\rm a}_4}$, are clearly visible in the {\it
PL} relations of both galaxies.

Sequence ${\rm a}_3$ corresponds to the ridge A of Wood \etal (1999)
while sequences ${\rm a}_2$ and ${\rm a}_1$ were merged in their ridge B
(Fig.~1).  Poorly populated ridge ${\rm a}_4$ has not been noticed in
earlier analyses.  Sequence ${\rm a}_3$ is the most numerous in the LMC
and SMC. Primary periods of almost half of the OSARGs of type ``a''
populate sequence ${\rm a}_3$ but for further 40\% of stars one of the
next dominant periods also corresponds to this sequence. About 20\% of
primary periods of OSARGs fall into the sequence ${\rm a}_2$, and only a
few percents of the variables have the primary periods in the sequences
${\rm a}_1$ and ${\rm a}_4$. The remaining OSARGs above the TRGB exhibit
the dominant periodicity in the sequence of long secondary periods
(LSP).

Fig.~5 indicates that the {\it PL} relations for $W_I$ index of OSARGs
brighter than the TRGB are not linear. The slope becomes steeper for
brighter $W_I$ indices. It is worth noticing that the slope of the {\it
PL} relations is different in the LMC and SMC, namely it is steeper in
the LMC than in the SMC.

Fig.~6 presents the Petersen diagram of OSARG stars from the LMC located
above the TRGB, that is a diagram where the period ratio of two periods in
a multi-period object is plotted against logarithm of the longer
period. Period ratios of all combinations of up to four dominant periods
in each object were plotted in the top panel of Fig.~6.
\begin{figure}[htb]
\vspace{12.5cm}
%\centerline{\includegraphics[width=13cm]{fig6.ps}} 
\FigCap{Petersen diagram for OSARG stars brighter than the TRGB in
the LMC. The large panel present ratios of the four dominant periods of
all these stars. The small diagrams present objects pulsating with periods
located in the four sequences in the {\it PL} diagram (Fig.~5).}
\end{figure} 

Six smaller panels present the Petersen diagrams for objects that possess
periods in the four sequences, ${{\rm a}_1-{\rm a}_4}$, in the {\it PL}
diagram. For example, in the left-top panel the ratios of the periods for
objects that pulsate simultaneously with periodicities falling into
sequence ${\rm a}_1$ and ${\rm a}_2$ are presented. All combinations of
sequences are shown. One can easily find a characteristic period ratio of
the sequences and its dependence on period.

It can be noted from Fig.~6 that the considerable fraction of the points in
the Petersen diagram is located in the region of period ratios larger than
0.97. Such very closely spaced frequencies were also detected in other
types of pulsating stars, \eg RR~Lyr variables (Olech \etal 1999) or
Cepheids (Moskalik, Ko{\l}aczkowski and Mizerski 2003), and were
interpreted as an indication of non-radial oscillations. By analogy we
suppose that these objects also pulsate non-radially.

The phenomenon of close periodicities is very common in OSARGs. For about
35\% of OSARGs in the LMC and 30\% in the SMC we found two (or more) close
frequencies between five highest peaks in the periodograms. More detailed
analysis of the periodograms would probably significantly increase the
number of stars oscillating in non-radial modes.

In about 55\% of OSARGs long secondary periods (LSP, Houck 1963),
\ie per\-iods typically 5--15 times longer than primary periods,
were detected. A clump of points representing these periodicities is
clearly visible in the lower-right part of the Petersen diagram. For about
half of the stars exhibiting LSP, amplitudes of the long period variability
are larger than amplitudes of typical OSARG's pulsation modes -- see an
example in the bottom panel of Fig.~4. However, the behavior of other
periodicities in these stars, like their period ratios, is similar to other
OSARGs. Therefore we believe that these objects constitute a common group
with typical OSARG stars.

It is worth noticing that LSPs fall into distinct {\it PL} sequence in
Fig.~5, roughly consistent with the sequence D of Wood \etal (1999). The
origin of the LSP phenomenon is still unknown. Possible explanations
include ``strange'' modes of pulsation, rotation of spotted star, episodic
dust ejections, or eclipses of giant by a cloud of dust and gas surrounding
the orbiting companion. More details concerning LSP objects are presented
by Wood, Olivier and Kawaler (2004).

Finally, the Petersen diagrams made it possible to measure the typical
ratio of periods corresponding to different sequences, ${{\rm a}_1-{\rm
a}_4}$, in the {\it PL} diagram. We obtained the following average values:
\begin{eqnarray*}
P_2/P_1\approx0.69&\\ 
P_3/P_1\approx0.50&\quad(P_3/P_1=-0.13\cdot\log P_1+0.74)\\ 
P_4/P_2\approx0.39&\\ 
P_3/P_2\approx0.73&\quad(P_3/P_2=-0.13\cdot\log P_2+0.96)\\ 
%\end{eqnarray*}
%\begin{eqnarray*}
P_4/P_2\approx0.56&\\
P_4/P_3\approx0.76&\\
\end{eqnarray*} 
\vspace*{-12mm}

One can notice that in some cases the period ratio depends on the
period. Typically, the longer period, the smaller period ratio. In the case
of $P_3/P_1$ and $P_3/P_2$ we provide above linear approximations of these
relations.

\vspace*{0.5cm}
\Section{OSARGs below the TRGB}
The nature of pulsating giants below the TRGB is still not clear. Alves
\etal (1998) and Wood (2000) argued that all these stars are thermally
pulsing AGB stars, while Ita \etal (2002) suggested that substantial
fraction of these objects are the first ascent RGB stars. Kiss and
Bedding (2003) provided additional arguments supporting this hypothesis.
They noted that the {\it PL} relations of variables above and below the
TRGB show a relative shift of about ${\Delta\log{P}\approx0.05}$, what
is consistent with the evolutionary temperature difference between AGB
and RGB stars with luminosities around the TRGB.

Our analysis of multi-periodic variable red giants provides evidences
supporting the hypothesis that below the TRGB both: AGB and RGB pulsating
stars are observed.  Moreover, we attempted to identify samples of first-
and second-ascent red giants fainter than the TRGB.

\begin{figure}[htb]
\vspace{12.9cm}
%\centerline{\includegraphics[width=12.5cm]{fig7.ps}} 
\FigCap{Period--$W_I$ diagrams for OSARG stars fainter than the TRGB
in the LMC (upper panel) and SMC (lower panel). Up to four dominant
periods for each star are plotted.}
\end{figure} 
OSARGs below the TRGB are much more numerous than variables above the
TRGB. Our sample consists of about 13\,400 stars fainter than the TRGB in
the LMC and 2600 such objects in the SMC. Fig.~7 presents the
$\log{P}$--$W_I$ diagram, for this group of stars from the LMC and
SMC. Similarly to OSARGs brighter than the TRGB four clear {\it PL} ridges
labeled from ${\rm b}_1$ to ${\rm b}_4$, can be distinguished as well as
the sequence of LSPs. The sequences denoted as ${\rm R}_1$, ${\rm R}_2$ and
${\rm R}_3$ by Kiss and Bedding (2003) below the TRGB correspond to the
sequences ${\rm b}_1$, ${\rm b}_2$ and ${\rm b}_3$, respectively, while the
sequences ${\rm A}^-$ and ${\rm B}^-$ of Ita \etal (2004) correspond to the
sequences ${\rm b}_3$ and ${\rm b}_2$, respectively.

The $\log{P}$--$W_I$ relations of OSARGs below the TRGB (Fig.~7) are
approximately linear and parallel. The slope of the relation in the SMC is
significantly smaller than in the LMC. Using the least square method we
fitted the slopes of ${-5.21\pm0.05}$ and ${-4.15\pm0.05}$ for OSARGs of
type ``b'' in the LMC and SMC, respectively. In Table~1 the corresponding
zero points of the relation for the sequences ${{\rm b}_1-{\rm b}_4}$ in
the $\log{P}$--$W_I$ plane are listed for the LMC and SMC. For comparison,
the zero points of OSARGs brighter than the TRGB fitted with the same
slopes for objects with ${W_I>10}$ and ${W_I>11}$ for the LMC and SMC,
respectively, are also listed (for brighter stars the slopes become
steeper).

\MakeTable{c@{\hspace{10pt}}
c@{\hspace{10pt}} c@{\hspace{10pt}} c@{\hspace{10pt}} c@{\hspace{10pt}}}
{12.5cm}{Zero points of the $\log P - W_I$ relations}
{\hline 
\noalign{\vskip3pt}
\multicolumn{5}{c}{Red Giants brighter than the TRGB}\\
\noalign{\vskip3pt}
\hline
Sequence & ${\rm a}_1$ & ${\rm a}_2$ & ${\rm a}_3$ & ${\rm a}_4$ \\
\noalign{\vskip3pt}
\hline
\noalign{\vskip3pt}
LMC & 20.37 & 19.59 & 18.86 & 18.21\\
\noalign{\vskip3pt}
\hline
\noalign{\vskip3pt}
SMC & 19.26 & 18.59 & 18.01 & 17.48\\
\noalign{\vskip3pt}
\hline
\noalign{\vskip3pt}
\multicolumn{5}{c}{Red Giants fainter than the TRGB}\\
\noalign{\vskip3pt}
\hline
\noalign{\vskip3pt}
Sequence & ${\rm b}_1$ & ${\rm b}_2$ & ${\rm b}_3$ & ${\rm b}_4$ \\
\noalign{\vskip3pt}
\hline
\noalign{\vskip3pt}
LMC & 20.65 & 19.91 & 19.07 & --\\
\noalign{\vskip3pt}
\hline
\noalign{\vskip3pt}
SMC & 19.52 & 18.88 & 18.22 & --\\
\noalign{\vskip3pt}
\hline}

\begin{figure}[htb]
\vspace{10.5cm}
%\centerline{\includegraphics[width=12.5cm]{fig8.ps}}
\FigCap{Period--$W_I$ diagram for all OSARG stars in the LMC. Cyan dots
mark OSARGs brighter than the TRGB (AGB stars), black dots OSARGs fainter
than the TRGB (RGB sample). Magenta points show the objects with one of the
dominant periods belonging to the shortest period sequence ${\rm b}_4$, \ie
AGB stars fainter than the TRGB. Broken line is the linear fit to
the sequence ${\rm a}_1$.}
\end{figure} 
\begin{figure}[htb]
\vspace{11.5cm}
%\centerline{\includegraphics[width=12.8cm]{fig9.ps}} 
\FigCap{Distribution of $\log P$ after subtraction of the
${\rm a}_1$ sequence fit. {\it Panel~A}: OSARG stars brighter than the
TRGB; {\it panel~B}: All OSARG stars fainter than the TRGB; {\it panel~C}:
AGB sample of OSARG stars fainter than the TRGB; {\it panel~D}: RGB sample
of stars fainter than the TRGB. Vertical dashed lines mark position of
sequences ${\rm a}_1-{\rm a}_4$.}
\end{figure} 
In further analysis of properties of OSARGs we limited ourselves to the
LMC red giants only, as this sample is the most numerous and the data
are most accurate. Fig.~8 presents the $\log{P}$--$W_I$ diagram for all
OSARG objects from Figs.~5 and 7 plotted together. The stars brighter
than the TRGB are marked with cyan dots. It is clear from Fig.~8 that
the ridges of these stars, ${{\rm a}_1-{\rm a}_3}$, are shifted in
$\log{P}$ relative to the corresponding ridges of stars fainter than the
TRGB: ${{\rm b}_1-{\rm b}_3}$, as noted by Kiss and Bedding (2003).

The location of the sequence indexed by 4, that went unnoticed in Kiss
and Bedding (2003) and Ita \etal (2004), indicates, however, that the
ridge of stars fainter than the TRGB, ${\rm b}_4$, is a straight
extension of the sequence ${\rm a}_4$ of OSARGs brighter than the TRGB,
\ie AGB stars. Fig.~9 shows the distribution of $\log{P}$ after
subtraction of the average linear fit for the sequence ${\rm a}_1$
(plotted with the broken line in Fig.~8)  for the red giants brighter
(panel~A) and fainter (panel~B) than the TRGB. Shifts and coincidence of
ridges indexed by 1--3 and 4, respectively, are again clearly seen.
Therefore we suspect that the stars fainter than the TRGB with one of
the dominant periods in the sequence ${\rm b}_4$ constitute the same
type stars as those in sequence ${\rm a}_4$, \ie the second ascent AGB
red giants.

To verify our hypothesis we closer examined the group of OSARGs fainter
than the TRGB with one of the periods located in the sequence ${\rm
b}_4$. The four dominant periods of these objects are marked by magenta
dots in the {\it PL} diagram (Fig.~8). The distribution of $\log{P}$ is
also presented in panel C of Fig.~9.

It can be immediately noticed from Figs.~8 and 9 that the dominant periods
of red giants from the sequence ${\rm b}_4$ form clear three additional
ridges in the {\it PL} diagram corresponding to those indexed by 1--3. What
more important these additional ridges are clearly shifted in $\log P$
compared to the ridges of the remaining OSARGs fainter than the TRGB (black
dots in Fig.~8). On the other hand they coincide perfectly with the ridges
${{\rm a}_1-{\rm a}_3}$ of giants brighter than the TRGB, \ie AGB stars and
are the extensions of the AGB stars ridges toward fainter objects.
Therefore we believe that the stars possessing the periods in the sequence
${\rm b}_4$ are indeed the pulsating red giants on the second ascent.

\begin{figure}[htb]
\vspace{12.5cm}
%\centerline{\includegraphics[width=12.9cm]{fig10.ps}}
\FigCap{Petersen diagram for OSARG stars fainter than the TRGB with
one of the four dominant periods in the sequence ${\rm b}_4$, \ie AGB
stars below the TRGB.}
\end{figure} 
\begin{figure}[htb]
\vspace{9cm}
%\centerline{\includegraphics[width=13.5cm]{fig11.ps}} 
\FigCap{Petersen diagram for OSARG stars fainter than the TRGB with one
of the four dominant periods in the sequence ${\rm b}_1-{\rm b}_3$, \ie
RGB sample.}
\end{figure} 
Fig.~10 shows the Petersen diagram for our group of pulsating AGB giants
fainter than the TRGB. The similarities with the diagram of AGB stars
brighter than the TRGB presented in Fig.~6 are striking. The period ratio
sequences presented in the small panels of Fig.~6 are long-period
continuations of the sequences of AGB red giants fainter than the TRGB
presented in Fig.~10.

The black dots in Fig.~8 mark the red giants fainter than the TRGB with the
dominant periods populating sequences ${{\rm b}_1-{\rm b}_3}$. We suppose
that the vast majority of these stars are the first ascent red giants, \ie
RGB giants (hereafter RGB sample). However, this sample can still be
contaminated to some extent by AGB stars. Although we already extracted a
group of AGB stars, based on the presence of ${\rm b}_4$ sequence
periodicity, it cannot be excluded that a number of AGB stars that do not
excite this mode of pulsation is still hidden among objects marked by black
dots. Unfortunately, the natural width of the {\it PL} sequences, accuracy of
observations and relatively similar distances in $\log P$ between sequences
indexed by 1--3 of the AGB and RGB stars make separation of additional
AGB stars impossible.

On the other hand the contamination of the RGB sample by hidden AGB stars
cannot be large. Panel D of Fig.~9 shows the distribution of $\log P$ of
the RGB sample. The expected maxima of the distribution of $\log P$ of AGB
stars (marked by vertical lines) are considerably shifted and fall between
the maxima of the RGB sample. If the number of AGB stars in the RGB sample
were considerable then the minima would be filled and much less pronounced
than seen in panel D of Fig.~9. Therefore, we conclude that our RGB sample
indeed constitutes in the statistical sense the RGB giants group.

Fig.~11 presents the Petersen diagram of the RGB sample. Similarly to stars
located above the TRGB, non-radial pulsations (period ratio close to 1.0)
and LSPs (period ratio smaller than 0.2) can be distinguished in Fig.~11 in
addition to the clear sequences corresponding to the period ratio of all
combinations of periods of sequences ${{\rm b}_1-{\rm b}_3}$. {\it PL}
relation of the LSP is a continuation of the LSP sequence of stars located
above the TRGB (\cf Figs.~5 and 7), and overlaps with the sequence D of
Wood \etal (1999). Both  period ratios -- close to 1.0 and LSP -- are also
present in the Petersen diagram of the AGB stars fainter than the TRGB
(Fig.~10).

A new feature clearly seen in the Petersen diagram in Fig.~11 are the
sequences corresponding to the period ratios of about 0.9 and 0.95. These
ratios are only observed among objects possessing pulsations in the sequence
${\rm b}_3$. The periodicities responsible for these two features form
additional sequences in the {\it PL} diagram merged with the ridge of ${\rm
b}_3$ in Figs.~7 and 8 and, thus, additionally widening it compared to
other ridges.

About 2000 stars possessing the period ratios between 0.88 and 0.92, and
about 1200 objects with the period ratio in the range 0.93--0.97 were
found. It is not clear whether the period ratios of about 0.9 and 0.95
correspond to radial pulsations or they represent non-radial
oscillations. Models of pulsating giants by Wood \etal (1999) permit such
very close radial modes of pulsations. It should be noted that the weak
signatures of these two period ratios are also seen in the Petersen diagram
of the AGB sample fainter that the TRGB (Fig.~10). It can indicate that
either the latter sample is still somewhat contaminated by RGB stars, or
that the pulsations responsible for these period ratios can be excited in
both types of giants.

\vspace*{0.5cm}
\Section{Discussion}
In this paper we presented large sample of pulsating red giants detected
during the OGLE-II survey in the Magellanic Clouds. We showed that the
sample can be divided into two groups with different pulsating
properties. We analyzed here in detail the group of small amplitude
pulsating red giants -- OSARGs.

We showed that the pulsational properties of the first and second ascent
OSARGs are slightly different what made it possible to show empirically
that pulsating red giants fainter than the TRGB are a mixture of RGB and
AGB objects and to select and extract subsamples of pulsating RGB and AGB
stars fainter than the TRGB. Follow up observations of these objects could
shed a~new light on the differences in mechanisms of pulsation of both
groups.

\begin{figure}[htb]
\vspace{9cm}
%\centerline{\includegraphics[width=13.5cm]{fig12.ps}} 
\FigCap{Petersen diagram for OSARG stars in  the SMC. Cyan dots mark
OSARG stars brighter than the TRGB, black points mark RGB sample OSARGs and
magenta dots OSARGs fainter than the TRGB with one of the four dominant
periods in the sequence ${\rm b}_4$, \ie AGB stars.}
\end{figure} 
Knowing the properties of OSARGs in the LMC, the natural question is
whether the same properties of small amplitude pulsating red giants are
shared by those in other stellar systems. The most obvious candidate for
the test is the SMC. Fig.~12 presents the Petersen diagram for the SMC
OSARGs. Cyan dots indicate stars brighter than the TRGB while black dots
those fainter than the TRGB. Additionally, the giants from the sequence
${\rm b}_4$ are marked by the magenta dots.

It is clear that the Petersen diagram of the SMC red giants closely
resembles that of the LMC objects. The giants from the sequence ${\rm b}_4$
form characteristic period ratio sequences coinciding with the similar
sequences of type ``a'' OSARGs, \ie AGB stars. Therefore, similarly to the
LMC we interpret those stars as AGB giants fainter than the TRGB. The
sample of these stars in the OGLE-II fields is, however, relatively
small. This might be related with different metallicity of both Magellanic
Clouds. Certainly much larger samples of these stars will become available
when OGLE-III data covering entire SMC are analyzed allowing more precise
studies of SMC OSARGs. It is also worth noticing that the slopes of {\it
PL} relations in the SMC are significantly different making these stars of
little use as standard candles.

Another important stellar system to test properties of OSARGs is the
Galactic bulge where Wray \etal (2004) discovered thousands of such objects
based on the OGLE-II photometry. Unfortunately, the Galactic bulge is much
worse region for analyzing the variable red giants using {\it PL} relations
because large and likely non-standard extinction (Udalski 2003b, Sumi 2004)
makes it difficult to obtain accurate intrinsic magnitudes of stars even in
the IR bands. Another factor diluting the {\it PL} diagrams is a large
geometrical depth of the Galactic bulge, \ie different distances to the
Galactic bulge giants.

Fortunately the periods and amplitudes of variable giants are free of these
problems. Therefore the Petersen diagram and period-amplitude diagram can
be directly compared with similar ones for the MC objects. It should be,
however, noted that the Galactic bulge sample was selected based on data
with much shorter time-span (only three years), so that considerable
fraction of periods in the Wray \etal (2004) list can be spurious or
non-stable. Therefore, one can expect more noise in the Galactic bulge
diagrams. One can also expect that because of the distance modulus smaller
by about 4~mag than that of the LMC the vast majority of red giants in the
OGLE-II Galactic bulge sample are stars fainter than the TRGB. The brighter
objects would be saturated in the OGLE-II images.

Fig.~13 presents the Petersen diagram for the Galactic bulge red giants
from the Wray \etal (2004) sample. All possible period ratios of the
dominant periods of these stars listed by Wray \etal (2004) are
plotted. As expected the diagram is much more noisy than the corresponding
diagram for the LMC OSARG stars fainter than the TRGB -- Figs.~10 and
11. Nevertheless, one can immediately notice striking similarities. The
characteristic period ratios in both diagrams are identical.

\begin{figure}[htb]
\vspace{8.3cm}
%\centerline{\includegraphics[width=13.5cm]{fig13.ps}}
\FigCap{Petersen diagram for OSARG stars in the Galactic bulge (Wray
\etal 2004). Magenta dots show position of OSARG stars with period
ratios corresponding to the AGB OSARG objects fainter than the TRGB in the
Magellanic Clouds. Black dots mark the remaining objects, most likely RGB
stars.}
\end{figure} 
Because the giants from sequence ${\rm b}_4$ cannot be extracted in the
Galactic bulge with the {\it PL} diagram as in the Magellanic Cloud cases
we selected them using the characteristic period ratio between sequences
${\rm b}_4$ and ${\rm b}_3$ equal to 0.76. This sequence is clearly seen in
Fig.~13. Period ratios of all dominant periods of these objects are marked
by magenta dots in Fig.~13.

It is striking in Fig.~13 that the selected objects populate the same
sequences as the LMC OSARGs from sequence ${\rm b}_4$, \ie the stars
interpreted by us as AGB stars fainter than the TRGB (Fig.~10). The
remaining giants populate sequences very similar to those of our RGB sample
of LMC giants (Fig.~11). Thus, it seems justified to conclude that the
small amplitude pulsating red giants in the Galactic bulge are also a
mixture of both -- AGB and RGB red giants and pulsational properties of
these stars in the Magellanic Clouds and the Galactic bulge are very
similar. Further comparison of properties of samples of these two class of
giants from different environments could shed the light on the dependence
of red giant pulsation mechanism on metallicity which is different in the
Magellanic Clouds and Galactic bulge.

It is also possible to explain the duality in the period--amplitude
diagram for small amplitude red giants discovered by Wray \etal (2004)
in the Galactic bulge. Group A of Wray's \etal giants  (variables with
lower amplitudes and shorter periods) consists mainly of the most
numerous class of OSARGs -- stars in the sequence ${\rm b}_3$. Variables
in the sequence ${\rm b}_2$ and stars brighter than the TRGB form group
B of Wray's \etal giants in the $\log{P}$--$\log{A}$ diagram.

\Acknow{We would like to thank Dr.\ W.\ Dziembowski for careful reading
of the manuscript, many important comments and suggestions. The paper
was partly supported by the Polish KBN grant 2P03D02124 to A.~Udalski.
Partial support to the OGLE project was provided with the NSF grant
AST-0204908 and NASA grant NAG5-12212 to B.~Paczy\'nski.}

\end{document}